\documentclass[preprint,aps,superscriptaddress,showpacs]{revtex4}
\usepackage{amssymb}
\usepackage{amsmath}
\usepackage{graphicx}
\usepackage[normalem]{ulem}
\usepackage[dvips]{color}
\usepackage{array,longtable,multirow}
\begin{document}

\title{Systematic Law for Half-lives of  Double $\beta$-decay with  Two Neutrinos }

\author{Yuejiao Ren}
\affiliation{Department of Physics, Nanjing University, Nanjing
210093, China}
\author{Zhongzhou Ren\footnote{Corresponding author, zren$@$nju.edu.cn}}
\affiliation{Department of Physics, Nanjing University, Nanjing
210093, China}
\affiliation{Center of Theoretical Nuclear Physics,
National
 Laboratory of Heavy-Ion Accelerator, Lanzhou 730000, China}

\begin{abstract}
Nuclear double $\beta$-decay with two neutrinos is a rare and
important process for natural radioactivity of  unstable nuclei. The
experimental data of nuclear double $\beta^{-}$-decay with two
neutrinos are analyzed and a systematic law to calculate the
half-lives of this rare process is proposed. It is the first
analytical and simple formula for double $\beta$-decay half-lives
where the leading effect from both the Coulomb potential and nuclear
structure is included.  The systematic law shows that the logarithms
of the half-lives are inversely proportional to the decay energies
for the ground state transitions between parent nuclei and daughter
nuclei. The calculated half-lives are  in agreement with the
experimental data of all known eleven nuclei with an average factor
of 3.06. The half-lives of other possible double $\beta$-decay
candidates with two neutrinos are predicted and these can be useful
for future experiments. The law, without introducing any extra
adjustment,  is also generalized to the calculations of double
$\beta $-decay half-lives from the ground states of parent nuclei to
the first $0^+$ excited states of daughter nuclei and the calculated
half-lives agree very well with the available data. Some calculated
half-lives are the first theoretical results  of double $\beta
$-decay half-lives from the ground states of parent nuclei to the
first $0^+$ excited states of daughter nuclei. The similarity and
difference between the law of $\alpha$-decay and that of double
$\beta^{-}$-decay are also analyzed and discussed.

\end{abstract}


\maketitle

The discovery of natural radioactivity by Becquerel changed our
views on the structure of matters and promoted the development of
modern physics and modern chemistry. Rutherford identified that
there are three kinds of natural radioactivity and named them as
$\alpha$-, $\beta$- and $\gamma$-rays \cite{li1,pre,sha,wa1}. The
structure of an atom is clear when Rutherford proposed the existence
of a nucleus in an atom based on $\alpha$-scattering experiments
with $\alpha$-particles from natural radioactivity.  Researches on
the problem of energy conservation in a  $\beta$-decay process lead
to Pauli's suggestion of the existence of a new particle, neutrino.
Based on this idea, Fermi proposed the basic formulation of the weak
interaction for the description of $\beta$-decay of a nucleus. In
1956, Lee and Yang proposed that parity cannot conserve for a weak
process such as $\beta$-decay \cite{lee}. Wu and her collaborators
carried out the $\beta$-decay experiment with polarized $^{60}Co$
nuclei and observed that parity symmetry was broken \cite{wu}. The
vector and axial-vector theory (V-A) of the weak interaction for
four fermions was founded in 1958 \cite{fey,sud} and is still widely
used for the calculations of $\beta$-decay half-lives and double
$\beta$-decay half-lives \cite{hax,eji,Int}. The unification of weak
and electromagnetic interactions is reached with the discovery of
intermediate bosons and the standard model is well founded with the
experimental confirmation of the Higgs particle. Currently new
progress on physics processes with neutrinos or without neutrinos is
being made. Some  important processes with neutrinos are the
measurement of $\theta_{13}$ in neutrino oscillations and the weak
processes such as the neutrino-induced reaction and the double
$\beta$-decay with neutrinos.  An important process without
neutrinos is the search of the nuclear double beta decay  without
emission of neutrinos. In this article we focus on the research of
half-lives of double $\beta$-decay with two neutrinos.

  There are many experimental and theoretical researches on double
 $\beta$-decay with two neutrinos and without neutrinos
\cite{hax,eji,ack,gan,arn,bar,saa,aud1,aud2,kla,rad,avi,ago,fae1,fae2,cau,suh,suo,vog,ver}.
  A list of available experimental data on decay energies and half-lives
  can be found in recent review articles \cite{bar,saa} and in
  nuclear data tables \cite{aud1,aud2}. Some recent experimental results on the
  half-lives of $^{130}Te$ are published in reference \cite{arn} and those of
  $^{136}Xe$ are published in
  references \cite{ack,gan}.  Very recent data of
  $^{76}Ge$ are reported in reference \cite{ago}. As
  experimental data are accumulated for 11 double
  $\beta$-decay nuclei with neutrinos, to make a
   systematic analysis on them and to find a systematic  law
   of data are useful for future researches.

      Theoretically there are two kinds of calculations
   on double $\beta$-decay half-lives and nuclear matrix elements
  \cite{hax,fae1,avi}. One is based on the nuclear shell model with
  an effective interaction from a renormalized g-matrix obtained with the Bonn
  potential \cite{hax,fae1,avi,cau}.
  Another is based on the  quasi-particle
  random-phase approximation
  (QRPA) or self-consistent renormalized
  random-phase approximation (RQRPA) \cite{hax,fae1,avi} where
   the collective particle-hole excitation of nuclei
   is included and the intermediate-nucleus states are represented as a set of
   phonon states  \cite{hax,fae1,avi}.
   A recent review can be found in reference \cite{avi}.
   These calculations are very successful for the description of
   the double $\beta$-decay half-lives and nuclear matrix elements
   due to the much effort of theoretical physicists.
They are in general very complicated, and require both sophisticated
computer codes and experiences in the specific problem to generate
results which can be compared with those obtained in the laboratory.
And in many cases different models generate different predictions,
which can be no easily reconciled. In the cases of the two neutrino
double $\beta $-decay, after decades of research there are plenty of
theoretical half-lives for many possible candidates based on the
calculations with the nuclear shell model or with
  the  various quasi-particle
  random-phase approximation. Although plenty of theoretical half-lives
  were reported for many nuclei, some recent experimental data
  are not covered by the calculations  \cite{saa}. For example, it is observed
  there is the double $\beta$-decay from the ground state of $^{150}Nd$
  to the first excited $0^+$ state of its daughter nucleus and there
  is no theoretical half-life on this \cite{saa}.  Therefore
  further calculation on double $\beta $-decay half-lives is needed.  Because there
  are plenty of calculations  based on the
 the nuclear shell model and
  the  various quasi-particle
  random-phase approximation, we do not carry out similar
  calculations  to them and we try a different way to investigate the
  double $\beta $-decay half-lives in this article. It is believed
  that the different approach to  double $\beta $-decay is useful for
  future development of the field and is also useful for future
  experimental research. It is widely accepted  that only experimental data
  themselves can test the reliability of  the prediction of the
  different approaches in physics.

  At first we make a systematic analysis on the available data
  of double $\beta$-decay to the ground state of daughter nuclei.
 Up to now there are 11 nuclei which are observed to have double
    $\beta$-decay with two neutrinos and their decay energies and half-lives
    are listed in Table 1.
    In Table 1, the first column denotes the parent nucleus and the
second column represents the experimental double $\beta$-decay
half-life for the ground-state transition between the parent nucleus
and the daughter nucleus. The experimental  double $\beta$-decay
half-lives are mainly presented  in the references
\cite{bar,saa,aud1} and here the data are  from  reference
\cite{saa}. We also list the logarithm of the half-life in column 3.
Column 4 is the experimental double $\beta$-decay energy of the
nucleus from the ground state of the parent nucleus to the ground
state of daughter nucleus where the
 data are from the nuclear mass table by Audi et al.
 and Wang et al. \cite{aud1,aud2}.
 The units of half-lives are Ey ($10^{18}$ years).
The fifth column represents the square root of the multiplication
between the logarithm of experimental half-life and the decay
energy.

\begin{table}[htb]
\small \centering \caption {The experimental data of double
$\beta$-decay half-lives (T$_{1/2}(expt.)$) and decay energies
($Q_{2\beta}$) of 11 even-even nuclei. In order to analyze the law
of the data we also list the logarithm of double $\beta$-decay
half-lives and the square root of the multiplication between the
logarithm of experimental half-life and the decay energy in columns
3 and 5, respectively. The units of the half-lives are Ey ($10^{18}$
years). The experimental half-lives and decay energies   in the
table are from  references \cite{saa} and \cite{aud2},
respectively.} \label{tab1}
\renewcommand{\tabcolsep}{2mm}
\begin{tabular}{ccccc}
\hline \hline Nuclei
        &T$_{1/2}(expt.)$(Ey)
        &lgT$_{1/2}(expt.)$
         &$Q_{2\beta}$(MeV)
        &$\sqrt{lgT_{1/2}(expt.)}\,\sqrt{Q_{2\beta} (expt.)}$)\\
\hline
$^{48}$Ca & $44^{+6}_{-5}$& 1.643& 4.267 & 2.649 \\
$^{76}$Ge & $1840^{+140}_{-100}$& 3.265&2.039 &2.580 \\
$^{82}$Se & $92^{+7}_{-7}$& 1.964& 2.996 & 2.424 \\
$^{96}$Zr & $23.5\pm{0.21}$ &1.371 & 3.349 & 2.143 \\
$^{100}$Mo &  $7.1\pm{0.4}$& 0.851& 3.034 & 1.606  \\
$^{116}$Cd &$28\pm{2}$& 1.447 & 2.813& 2.017 \\
$^{128}$Te & $(1.9\pm{0.4})\times 10^6$ & 6.279&0.8665 & 2.333 \\
$^{130}$Te & $700\pm{140}$  &2.845 & 2.528 & 2.682 \\
$^{136}$Xe & $2300\pm{120}$  & 3.362 &2.458 &  2.876 \\
$^{150}$Nd & $9.11\pm{0.68}$  & 0.960 & 3.371 &  1.799 \\
$^{238}$U & $2000\pm{600}$& 3.301 & 1.144  &  1.944 \\
\hline \hline
\end{tabular}
\end{table}

It is seen from Table 1 that the shortest half-life is $7.1\ Ey$ for
$^{100}Mo$  and the longest is  $1.9\times 10^6 \ Ey$ for
$^{128}Te$. The half-lives vary  from $7.1\ Ey$ ($^{100}Mo$ ) to
$1.9\times 10^6 \ Ey$ ($^{128}Te$) although the change of the decay
energies  is less than four times ( from 3.034 MeV for $^{100}Mo$ to
0.8665 MeV for $^{128}Te$). This shows that the half-life is very
sensitive to the decay energy and the relationship between them may
be very complex by a glance of the data. In order to describe
quantitatively the sensitivity and to find a law among the data, we
introduce the logarithm of the half-life and also make a square root
for the multiplication between the  logarithm of the half-life and
the decay energy. The square root of the  multiplication is listed
in the last column of Table 1.  It is clearly seen from the last
column that the variation of the square root of the multiplication
is in a very narrow range from the minimum 1.606 to the maximum
2.876. This fact strongly suggests that a constant value can be a
good approximation for the multiplication of the logarithm of the
half-lives and decay energies. The physics behind this will be
discussed later in this article. This is the starting point of our
researches and we simply write it in the following mathematical
equation between   the logarithm of half-life and the decay energy,
\begin{eqnarray}\label{eq1}
lg\,T_{1/2}(Ey)\,=\,a\,/{Q_{2\beta} (MeV)}
\end{eqnarray}
Here $a$ is a constant and its value is to be determined. We will
discuss the meaning of this constant later and will point out  that
the constant $a$ is from the universal properties of the weak
interaction for the decay process of different nuclei.

  For a better description of the double $\beta$-decay half-lives, the effects
  from the Coulomb potential and from nuclear structure should be taken into account.
It is well known from the $\beta$-decay theory \cite{pre,sha,wa1}
that the effect from the Coulomb potential can be derived from
quantum mechanics and  its expression in the nonrelativistic case is
given in some textbooks \cite{pre,sha,wa1},
\begin{eqnarray}\label{eq2}
\rho (Z, \epsilon)\,=\, 2\,\pi \eta /(1-e^{-2\,\pi \eta })
\end{eqnarray}
where $\eta=\,(Z/137)\,(\epsilon/cp)$ and $\epsilon$ is the energy
of electron and $p$ is the magnitude of the electron momentum.

Usually the correction of the Coulomb potential should  multiply the
square of the matrix element to obtain the probability of
$\beta$-decay  \cite{pre,sha,wa1}.  This correction factor is close
to unity for light nuclei but it has an evident effect for heavy
nuclei. In the usual $\beta $ decay or the double $\beta$-decay of
this article, the speed of an electron is very close to the speed of
light ($v_e\approx 0.86-0.95\, c$) and therefore $(\epsilon/cp)$ is
close to unity. For the denominator of Eq.(2) it is also a good
approximation to choose it to be unity for heavy nuclei. Therefore
 we only keep the leading term of equation (2) for simplicity and
 a correction of the Coulomb potential for double
 $\beta$-decay half-lives  is approximately
  $-2\,lg((2\pi Z)/137)$
 to the numerator of equation (1) where Z is chosen to be the charge number
 of parent nuclei and this is also for the convenience of calculations.
  This correction is approximately zero
 for light double $\beta$-decay nuclei such as $^{48}Ca$ and it
 can give significant contribution for heavy nuclei such as
 $^{238}U$.

   Now let us  take the nuclear structure effect into account to
   some extent. We keep with the
   same spirit to the correction of the Coulomb potential and only
   include the leading effect from the strong interaction. For
   nuclear structure, the most important effect from the strong
   interaction of nucleons is the existence
   of magic numbers and this corresponds to the appearance of nuclear shell
   structure in nuclei.
   The nuclei with magic number are more stable than non-magic
   nuclei.
   For the  numerical calculations of both shell model and mean-field
   model,
   the first step is the choice of the  number of valence
   nucleons. The choice of the  number of valence
   nucleons is with reference  to magic numbers. The number of valence neutrons
   of a nucleus is zero if its neutron number corresponds to a magic number.
   The  valence neutron  number of a nucleus  is not zero if
   its neutron number does not correspond to a magic number.
   The parameters of the effective mean-field potential
   in a self-consistent mean-field model are often obtained with the fitting of ground
   state properties of several magic nuclei such as $^{16}O$, $^{40,48}Ca$,
   and $^{208}Pb$ and the residual interaction such as the pairing force
   is included for valence nucleons when it is used for researches of open shell nuclei.
   For  the double $\beta$-decay process it involves a change of two
   neutrons into two protons for nuclei with the same nucleon number
   (A). It can be observed if other decay mode such as a
   single $\beta$-decay is forbidden in energy.  Based on this we simulate the
   leading effect of nuclear structure (i.e. shell effect)  by introducing an addition
   quantity $S$ to the numerator of equation (1).   $S=2$ when
   the neutron number of a parent nucleus is a magic number
   and $S=0$ when the neutron
   number is non-magic. Whether this choice is suitable can be tested
   by the numerical results of this article.
   In this way we do not touch the details of
   the complex nuclear potential and we also avoid solving the very
   complicated nuclear many-body problem.  So a new systematic law of double
   $\beta$-decay half-lives is proposed to be
\begin{eqnarray}\label{eq3}
lg\,T_{1/2} (Ey)\,=\,(a\,-2\,lg(2\,\pi
\,Z/137)\,+\,S\,)\,/\,{Q_{2\beta} (MeV)}
\end{eqnarray}
where the constant $a$ is obtained by fitting the experimental data
of Table 1 and its value is $a=5.843$. We will point out later that
the physical meaning of $a$ is related to the square of the strength
of the weak interaction which leads to the instability of a nucleus.
$Z$ is the charge number of the parent nucleus. $S=2$ when the
neutron number of parent nuclei is a magic number and \,$S=0$ when
the neutron number of parent nuclei is not a magic number.  The
number of parameters of equation (3) is two if both $a$ and $S$ are
considered to be adjusting parameters.  The number of parameters of
equation (3) is one  if only  $a$ is considered to be an adjusting
parameter and if $S$ is accepted to be from the shell effect (in the
shell model and in the mean-field model people naturally choose the
valence number with reference of magic numbers).  In a word, the
number of adjusting parameters in equation (3) is the least as
compared to a model with an effective potential because one needs
two parameters (depth and force range) to define a central potential
and a third parameter to define the strength of the spin-orbit
potential.  Because the equation (3) is very clear and simple, any
physicists including experimental physicists can use it to calculate
the half-lives of double $\beta$-decay with inputs of the decay
energies.  A pocket calculator with logarithm function is  enough to
get numerical results.

  The numerical results from equation (3) are drawn in Figure 1 and also
  listed in Table 2,
  together with experimental half-lives and decay energies. In
  Fig.1, the X-axis is the nucleon number and the Y-axis is the
  logarithm of the double $\beta$-decay half-lives. Experimental
  data and calculated results with equation (3) are denoted with
  different symbols in the figure.  It is seen that the calculated
  results are close to experimental data and reasonable agreement
  is achieved for eleven nuclei which are observed to have double
  $\beta$-decay with two neutrinos. We  make a detailed
  discussion on the calculated results of Table 2.
  In Table 2  columns 1-4 are experimental data and they have the same meaning
  as columns 1-4 of Table 1. Column 5 is   the logarithm of the calculated
  double $\beta$-decay half-lives with equation (3). The last column
  is the logarithm of the ratio between experimental double
  $\beta$-decay half-lives and calculated ones and it shows the
  deviation between experimental half-lives and calculated ones.
  When the deviation is less than 0.301, the agreement between
  the experimental half-life and the calculated one is within a factor of
  two (lg2=0.301) and this corresponds to the cases for $^{48}Ca$, $^{82}Se$,
  $^{96}Zr$ .  When the deviation is less than 0.477, the agreement between
  the experimental half-life and the calculated one is within a factor of
  three (lg3=0.477) and this is the case for
   $^{116}Cd$ and $^{128}Te$.
  It is seen from Table 2 that experimental half-lives can be
  reproduced by calculations within a factor of four for eight nuclei
  (lg4=0.602) (such as  the case of $^{76}Ge$, $^{136}Xe$ and  $^{150}Nd$).
    The deviation between experimental half-lives  and the calculated
  results is beyond a factor of four for three nuclei
  $^{100}Mo$,$^{130}Te$, and $^{238}U$. For $^{238}U$ the agreement
  for half-lives is approximately a factor of 5.2 (lg5.2=0.716).
 For $^{100}Mo$ the calculation can reproduce its
  experimental half-life within a factor of eight
  (lg8=0.903) and for $^{130}Te$ the calculation can reproduce that within a factor of
  seven (lg7=0.845).  In order to see the total agreement for eleven
  nuclei we define an average deviation
  $\Delta=\{\Sigma_{i=1}^{i=11}\
  \left|lg(T_{1/2}(expt.)/T_{1/2}(theo.))\right| \}/11=0.486$. This
  corresponds  to a factor of 3.06 between experimental half-lives
  and calculated ones (lg3.06=0.486).  This agreement between double $\beta$-decay
  half-lives and calculations with equation (3) is good compared
  with the calculations of
single $\beta$-decay half-lives \cite{ni6,ni7,li6,zha,mol} and
${\alpha }$-decay half-lives of unstable nuclei
\cite{ni1,ni2,ren,ren3}. Especially the calculations of this article
are with the minimum number of adjusting parameters and cover a wide
range of nuclear mass from $A=48$ ($ ^{48}Ca$) to $A=238$ ($
^{238}U$).  This means it can be used to predict the double
$\beta$-decay half-lives of other nuclei in nuclide charts.

\begin{table}[htb]
\small \centering \caption {The logarithm of double $\beta$-decay
half-lives of even-even  isotopes calculated with new  law
($lgT_{theo.}$) and the corresponding experimental ones
($lgT_{expt.}$). The units of the half-lives are Ey ($10^{18}$
years). The experimental decay energies of nuclei ($Q_{2\beta}$
(MeV)) are also listed in the table where the decay energies are
from the nuclear mass table  \cite{aud2}. The calculated half-lives
are  in agreement with the experimental data of all known eleven
nuclei with an average factor of 3.06. The maximum deviation between
theoretical  half-life and experimental one is a factor of 7.709 and
this corresponds to the case of $^{100}Mo$.} \label{tab2}
\renewcommand{\tabcolsep}{2mm}
\begin{tabular}{cccccc}
\hline \hline Nuclei
        &T$_{1/2}(expt.)$(Ey)
        &$lgT_{1/2}(expt.)$
         &$Q_{2\beta}$(MeV)
        &$lgT_{1/2}(theo.)$&$lg(T_{1/2}(expt.)/T_{1/2}(theo.))$\\
\hline
$^{48}$Ca & $44^{+6}_{-5}$& 1.643& 4.267      & 1.856  &-0.213 \\
$^{76}$Ge & $1840^{+140}_{-100}$& 3.265&2.039 & 2.702 & +0.563 \\
$^{82}$Se & $92^{+7}_{-7}$& 1.964& 2.996      & 1.822  & +0.142 \\
$^{96}$Zr & $23.5\pm{0.21}$ &1.371 & 3.349    & 1.587  & -0.216 \\
$^{100}$Mo &  $7.1\pm{0.4}$& 0.851& 3.034     & 1.738   & -0.887 \\
$^{116}$Cd &$28\pm{2}$& 1.447 & 2.813         & 1.834  &  -0.378 \\
$^{128}$Te & $(1.9\pm{0.4})\times 10^6$ & 6.279&0.8665 & 5.872 &0.407 \\
$^{130}$Te & $700\pm{140}$  &2.845 & 2.528    & 2.013  & 0.832 \\
$^{136}$Xe & $2300\pm{120}$  & 3.362 &2.458   &  2.871 & 0.491     \\
$^{150}$Nd & $9.11\pm{0.68}$  & 0.960 & 3.371 &  1.473 & -0.513 \\
$^{238}$U & $2000\pm{600}$& 3.301 & 1.144     &  4.015 & -0.714 \\
\hline \hline
\end{tabular}
\end{table}

     In Table 3 we predict the double $\beta$-decay half-lives with two neutrinos
    for eleven nuclei. This corresponds to the decays between the
    ground states of parent nuclei and daughter nuclei and their mass numbers
    lie in a very wide range from $^{46}Ca$ to $^{244}Pu$. They are
    the good candidates to observe the  double $\beta$-decay  with two
    neutrinos. In Table 3, the first column denotes the parent nuclei
    and the second column is the experimental decay energies  where the
    experimental data are from the nuclear mass tables \cite{aud1,aud2}.
    The third column corresponds to the logarithm of calculated
    half-lives  with equation (3). We also list the double $\beta$-decay
    half-lives  in the last column and this is
    convenient to make a direct comparison with future double $\beta$-decay
    experiments. The decay energies of these nuclei vary
    approximately from 1 MeV to 2 MeV and their half-lives range
    approximately from $10^2$ Ey to $10^6$ Ey.  These will be tested
    by future measurements.

    Recently it is also observed that the double $\beta$-decay with
    two neutrinos can occur from the ground state of parent nuclei
    to the first $0^+$ excited state of daughter nuclei \cite{saa}.
    Based on the successful description of the double $\beta$-decay
    half-lives of ground-state transitions with equation (3),
    we now generalize the equation (3)
    to the calculation of double $\beta$-decay
    half-lives to the first $0^+$  excited state of daughter nuclei.
    We list the numerical results in Table 4, together
    with experimental decay energies and two experimental half-lives
    from the decay of $^{100}Mo$ and $^{150}Nd$ to the first  $0^+$
     excited state of the daughter nuclei \cite{saa}.  For the half-life
     from $^{100}Mo$, the calculated result is approximately the same as
     the experimental one. For that of  $^{150}Nd$, the calculated result agrees with
      experimental one within a factor of two ($lg2=0.301$). It is seen without
      introducing any extra adjustment in the law that very
     perfect agreement is reached for the available data when the law is extended
     to the decay  to the first $0^+$ excited state of daughter
     nuclei.  This confirms  the validity of the law.  According to the newest review
     article of  the double $\beta$-decay with
    two neutrinos \cite{saa}, no theoretical half-life is reported
    for the decay from  $^{150}Nd$ to the first  $0^+$ excited
     state of the daughter nucleus. Therefore the calculated half-life
     of $^{150}Nd$
     in Table 4 is the first theoretical result. Besides the experimental
     half-lives of $^{100}Mo$ and $^{150}Nd$, there are not definite half-lives
     from the measurement of  other nuclei  in Table 4  \cite{saa}. Our
     calculated results are above the low limit of the half-lives for
     other nuclei \cite{saa}.

\begin{table}[htb]
\small \centering \caption {The  double $\beta$-decay half-lives
($T_{1/2}(theo.)$) of even-even  isotopes calculated with new
systematic law and the corresponding logarithms
($lgT_{1/2}(theo.)$). The units of the half-lives are Ey ($10^{18}$
years). The experimental decay energies of nuclei ($Q_{2\beta}$
(MeV)) are also listed in the table where the decay energies are
from the nuclear mass table  \cite{aud2}.} \label{tab3}
\renewcommand{\tabcolsep}{2mm}
\begin{tabular}{cccc}
\hline \hline Nuclei
         &$Q_{2\beta}$(MeV)
        &$lgT_{1/2}(theo.)$&$T_{1/2}(theo.)$(Ey)\\
\hline
$^{46}$Ca  & 0.989  & 5.984   & $9.64\times 10^5$ \\
$^{86}$Kr  & 1.258  & 5.889   & $7.74\times 10^5$ \\
$^{94}$Zr  & 1.142  & 4.655   & $4.52\times 10^4$ \\
$^{104}$Ru & 1.301  & 4.023   & $1.05\times 10^4$ \\
$^{110}$Pd & 2.017  & 2.576   & $3.77\times 10^2$ \\
$^{148}$Nd & 1.928  & 2.575   & $3.76\times 10^2$ \\
$^{154}$Sm & 1.251  & 3.945   & $8.81\times 10^3$ \\
$^{160}$Gd & 1.731  & 2.835   & $6.84\times 10^2$ \\
$^{198}$Pt & 1.049  & 4.515   & $3.27\times 10^4$ \\
$^{124}$Sn & 2.291  & 2.236   & $1.72\times 10^2$ \\
$^{244}$Pu & 1.35   & 3.388   & $2.44\times 10^3$ \\
\hline \hline
\end{tabular}
\end{table}

\begin{table}[htb]
\small \centering \caption {This table corresponds to double
$\beta$-decay from the ground state of parent nuclei to the first
$0^+$ excited state of  daughter nuclei (denoted with a symbol $*$).
Column 1 represents the parent nuclei and column 2 denotes the
experimental double $\beta$-decay half-lives \cite{saa}. Our
calculated half-lives from the systematic law are listed in column
4. The units of the calculated half-lives in column 4 are Ey
($10^{18}$ years). The experimental decay energies of nuclei
($Q^*_{2\beta}$ (MeV)) are also listed in the table where the
experimental data  are from reference \cite{saa}. The calculated
results of $^{48}$Ca and $^{150}$Nd are the first theoretical
results according to Table 3 of the newest review article
\cite{saa}. The calculated results (Ey) from other groups are also
listed in the last two columns for comparison and these are also
taken from Table 3 of the review article \cite{saa}.} \label{tab4}
\renewcommand{\tabcolsep}{2mm}
\begin{tabular}{cccccc}
\hline \hline Nuclei
        &$T^*_{1/2}(expt.)$(Ey)
         &$Q^*_{2\beta}$(MeV)&$T^*_{1/2}(theo.)$(Ey)
         &$T^*_{1/2}(other1.)$(Ey)&$T^*_{1/2}(other2.)$(Ey)\\
\hline
$^{48}$Ca  &      & 1.275       &$1.63\times 10^6$   &  &  \\
$^{76}$Ge  &        &0.917      &  $1.02\times 10^6$
& $(7.5-310)\times 10^3$ \cite{s41,s66}&$4.5\times 10^3$ \cite{s67}\\
$^{82}$Se  &        & 1.506    &$4.21\times 10^3$
&$(1.5-3.3)\times 10^3$ \cite{s41,s66}&\\
$^{96}$Zr  &        & 2.203      &$2.59\times 10^2$
& $(24-27)\times 10^2$ \cite{s41,s66}&$38\times 10^2$ \cite{s67}\\
$^{100}$Mo & $5.9^{+0.8}_{-0.6}\times 10^2$   & 1.904
    &$5.89\times 10^2$ &$16\times 10^2$ \cite{s74} &$21\times 10^2$ \cite{s67}\\
$^{116}$Cd &       & 1.048         &$8.36\times10^4$
&$1.1\times 10^4$ \cite{s41,s66} &$0.11\times 10^4$ \cite{s67}\\
$^{130}$Te &       & 0.735      & $8.38\times 10^6$
&$(5.1-14)\times 10^4$ \cite{s41,s66,s76}&\\
$^{150}$Nd & $1.33^{+0.45}_{-0.26}\times 10^2$ & 2.627
  &$0.776\times 10^2$ & &\\
\hline \hline
\end{tabular}
\end{table}

  After we present numerical results for double $\beta$-decays to
  both the  ground state and to the first $0^{+}$ excited state of
  daughter nuclei, it is useful to discuss the physics behind the new law.
  For this purpose we compare the differences and common points
between $\alpha$-decay and double $\beta$-decay because both of them
are two important decay modes of unstable nuclei. It is  known that
the half-lives of  $\alpha$-decay can be calculated by the
Geiger-Nuttall law or the Viola-Seaborg formula with a few
parameters \cite{ni1,ren,ren3}. A new version of the Geiger-Nuttall
law is proposed \cite{ren} and it can well reproduce the
experimental half-lives of both $\alpha$-decay and cluster
radioactivity \cite{ni1,ren,ren3}. The new Geiger-Nuttall law
between $\alpha$-decay half-lives (in seconds) and $\alpha$-decay
energies (in MeV) of ground-state transitions of even-even nuclei is
\cite{ren}

\begin{eqnarray}\label{eq2}
lgT^{\alpha}_{1/2}(seconds)\,=\,a\,{\sqrt{\mu}}\,{Z_c}{Z_d}/{\sqrt{Q^{\alpha}}}
+\,b{\sqrt{\mu}}{\sqrt{Z_cZ_d}}+\,c\,+\,S\,
\end{eqnarray}
  In this equation the values of three parameters are $ a=0.39961,
b=-1.31008, c_{e-e}=-17.00698$ for even-even (e-e) nuclei
\cite{ren}. $T^{\alpha}_{1/2}$(seconds) is the half-life of
$\alpha$-decay and $Q^{\alpha}(MeV)$ is the corresponding decay
energy. $Z_c$ and $Z_d$ are the charge numbers of the cluster and
the daughter nucleus, respectively. $\mu=A_c\,A_d/(A_c+A_d)$ is the
reduced mass and $A_c$,$A_d$ are the mass numbers of the cluster and
daughter nucleus, respectively.  For $\alpha$-decay, $Z_c$=2 and
$A_c$=4. The three parameters, $a,b,c$, are obtained by fitting the
data of even-even nuclei with $Z\ge 84$ and $N\ge 128$ \cite{ni1}.
$S$ is a new quantum number  to mock up the shell effect of N=126 on
$\alpha$-decay half-lives. The value of $S$ for ground-state
transitions of even-even nuclei is: $S=0$ for $N\ge \,128$ and $S=1$
for $\ N\leq \,126$ \cite{ren}.

  When we compare the new Geiger-Nuttall law of $\alpha$-decay half-lives
  (equation (4)) with
  the new systematic law of double $\beta$-decay half-lives (equation
  (3)), we find they are very similar although they are governed by
  different interactions.  For a clear comparison, one can keep the
  first term of  both laws temporarily  because it is the leading term and the last
  two terms are the corrections to the leading term ( the constant
  $c$ in new Geiger-Nuttall law can be approximately
  eliminated if we change the units of the half-life
  from seconds to $10^{17}$ seconds). A common point between $\alpha$-decay
  and double $\beta$-decay  is that their half-lives are very
  sensitive to the decay energy and the equations on their half-lives look alike.
  The first term in new
  Geiger-Nuttall law (equation (4)) is dependent on both charge numbers and decay
  energies because the Coulomb repulsive potential leads to the
  appearance of $\alpha$-decay (a quantum tunneling effect) and
  the total effect from the Coulomb potential
  is related to the charge numbers (similarly the total effect of the strong
  interaction is also directly related to the nucleon number of a nucleus).
  For the systematic law of
   double $\beta$-decay half-lives (equation (3)), the first term is
   only dependent on the decay energy because the weak interaction is
   universal for natural decay processes
   and the total effect from the weak interaction is not
   very sensitive to the change of nucleon numbers (such as
   proton numbers).  This agrees with our knowledge of the weak
   interaction that the strength of the weak interaction of a free-neutron
   $\beta$-decay is approximately same as that of a $\mu$ decay
   although two decay systems are very different and the difference of their decay
   energies is approximately 100 times (of course the strength of a weak
   interaction can be different  if different generation
   of quarks or different generation of leptons in Standard Models are involved.)
     It is due to the difference of
   this total effect between the weak interaction and the Coulomb
   interaction that  $\beta$-decay occurs for the ground state of
   many unstable nuclei from very light ones (such as
   a decay from a neutron or from a triton) to heavy ones
   and $\alpha$-decay occurs for ground states of medium and heavy nuclei.
   Another important difference between the new systematic law and the
Geiger-Nuttall law is from the difference of the perturbation
approximation in quantum mechanics. The double $\beta$-decay is a
second-order process of the weak interaction  with the V-A
four-fermion theory where a single $\beta$-decay is forbidden in
many double $\beta$-decay nuclei. For the Geiger-Nuttall law, the
$\alpha$-decay is a first-order process of the electromagnetic
interaction and there are significant influences from the strong
interaction.  Before ending the discussion, we would like to  point
out that the right side of Eq.(3) can be dimensionless if one would
like to replace the decay energy $Q_{2\beta}$ by $Q_{2\beta
}/(2m_ec^2)$. In this case the accuracy of calculated half-lives
with the systematic law is almost kept due to
$2m_ec^2=1.022\,MeV\,\approx 1 \,MeV$.

\begin{figure}[htb]
\centering
\includegraphics[width=12cm]{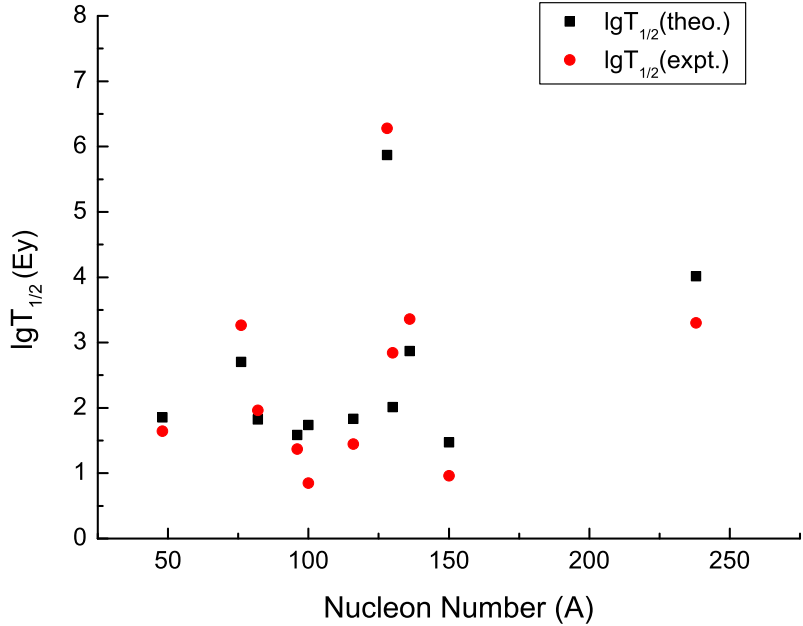}
\caption{ Logarithms of the experimental double $\beta$-decay
half-lives and theoretical ones for ground-state transitions of 11
even-even nuclei from $^{48}$Ca to $^{238}$U. The calculated
half-lives are in agreement with the experimental data of all known
eleven nuclei with an average factor of 3.06. The maximum deviation
between theoretical half-life and experimental one is a factor of
7.709 and this corresponds to the case of $^{100}Mo$.}
\end{figure}

In summary, the new law for the calculations of double $\beta$-decay
half-lives is proposed where the leading effects of the decay
energy, the Coulomb potential and the nuclear structure  are
naturally taken into account. This is the first analytical formula
 for  the half-lives of the complex double
$\beta$-decay with two neutrinos where  only two parameters are used
in the formula. By including these leading effects, the available
data of double $\beta$-decay half-lives of ground-state transitions
in even-even nuclei are reasonably reproduced. Without introducing
extra adjustment on the two parameters of the law, the law is
generalized to  the double $\beta$-decay between the ground state of
parent nuclei and the first   $0^{+}$ excited state of daughter
nuclei and perfect agreement between the calculated half-lives and
the data is reached. The existence of these terms is based on the
quantum theory for the microscopic
 description of the double $\beta$-decay process.
The universal behavior of the weak interaction manifests itself in
the formula of double $\beta$-decay half-lives by comparing  the
similarity and difference between the systematic law of double
$\beta$-decay half-lives and the famous Geiger-Nuttall law of
$\alpha$-decay.  The half-lives of the double $\beta$-decay
candidates with two neutrinos   are predicted and they are useful
for future experiments.

\

This work is supported by National Natural Science Foundation of
China (Grant Nos. 11035001, 10975072, 11120101005, 11175085), by the
Research Fund of Doctoral Point (RFDP), No. 20070284016 and by the
Priority Academic Program Development of Jiangsu Higher Education
Institutions.

\end{document}